\documentclass[%
 reprint,
 amsmath,amssymb,
 aip,
 jap,
]{revtex4-1}

\usepackage{graphicx}
\usepackage{dcolumn}
\usepackage{bm}

\begin{document}

\preprint{APS/123-QED}

\title[Sample title]{Quantum resistance standard accuracy close to the zero-dissipation state\\
}

\author{F. Schopfer, W. Poirier}
\email{wilfrid.poirier@lne.fr}
\affiliation{Laboratoire National de m\'{e}trologie et d'Essais (LNE), 29 avenue Roger Hennequin, 78197 Trappes, France}

\date{\today}

\begin{abstract}
 We report on a comparison of four GaAs/AlGaAs-based quantum resistance standards using an original technique adapted from the well-known Wheatstone bridge. This work shows that the quantized Hall resistance at Landau level filling factor $\nu=2$ can be reproducible with a relative uncertainty of $32\times 10^{-12}$ in the dissipationless limit of the quantum Hall effect regime. In the presence of a very small dissipation characterized by a mean macroscopic longitudinal resistivity $\overline{R_{xx}(B)}$ of a few $\mu\Omega$, the discrepancy $\Delta R_{\mathrm{H}}(B)$ measured on the Hall plateau between quantum Hall resistors turns out to follow the so-called resistivity rule $\overline{R_{xx}(B)}=\alpha B\times d(\Delta R_{\mathrm{H}}(B))/dB$. While the dissipation increases with the measurement current value, the coefficient $\alpha$ stays constant in the range investigated ($40-120~\mathrm{\mu A}$). This result enlightens the impact of the dissipation emergence in the two-dimensional electron gas on the Hall resistance quantization, which is of major interest for the resistance metrology. The quantum Hall effect is used to realize a universal resistance standard only linked to the electron charge \emph{e} and the Planck's constant \emph{h} and it is known to play a central role in the upcoming revised \emph{Syst\`eme International} of units. There are therefore fundamental and practical benefits in testing the reproducibility property of the quantum Hall effect with better and better accuracy.
\end{abstract}

\maketitle

\section{\label{sec:level1}Introduction}

The quantum Hall effect (QHE)\cite{Klitzing1980}, which manifests itself by the Hall resistance quantization in two-dimensional electron gas (2DEG) at $R_\mathrm{K}/i$ values in the non-dissipative transport limit ($i$ is an integer and $R_\mathrm{K}$ is the von Klitzing constant equal to $h/e^2$ in theory\cite{Yoshioka1998}, with $e$ the electron charge and $h$ is the Planck constant), led to a universal and reproducible representation of the unit ohm that can be currently maintained in national metrology institutes with a relative uncertainty \cite{Jeckelmann2001,Poirier2009} of $1 \times 10^{-9}$. This breakthrough also resulted from the advent of resistance bridges based on cryogenic current comparator (CCC)\cite{Harvey1972} using superconducting quantum interference devices (SQUID). The impact of the QHE in metrology was more recently enlarged by its implementation in the alternating current regime\cite{Ahlers2009} and by the development of quantized Hall resistances (QHR) arrays\cite{Poirier2002}. A similar breakthrough occurred in the field of voltage unit following the discovery of the Josephson effect (JE)\cite{Jeanneret2009}. The major asset of the ohm representation by the QHE is its excellent reproducibility resulting from its link to the fundamental constants of physics. Practically, quantum standards must be independent of the experimental implementation, material nature and sample properties, provided that some practical quantization criteria are fulfilled\cite{Delahaye2003}. Among those is the absence of dissipation in the 2DEG.
\begin{figure}[h!]
\begin{center}
\includegraphics[width=8.5cm]{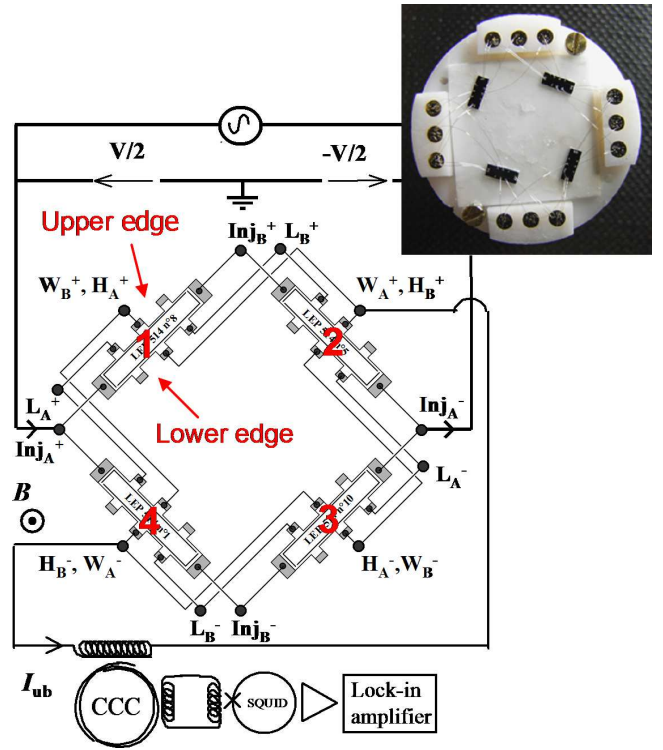}
\caption{Drawing of the experimental setup including the quantum Hall Wheatstone bridge made of four Hall bar samples (photography) and the CCC measuring the unbalance current.}\label{fig1}
\end{center}
\end{figure}
 As emphasized by the Landauer-B\"uttiker theory of the QHE\cite{Buttiker1988}, the perfect quantization of the Hall resistance in the QHE regime relies on the absence of any backscattering of electrons. In practice, this means the measurement of a very low value of the longitudinal resistance. In Hall bar samples of metrological quality, the Hall resistance was observed to deviate linearly from quantized value as a function of the minimum longitudinal resistance\cite{Cage1984,Jeckelmann2001}. Even if explanations have been proposed for this observation\cite{Vanderwel1988, DominguezPhd1988, Granot2000}, the relationship between the two resistivity components remains to be further investigated as a function of magnetic field very close to the zero-dissipation state.

 It remains a continuous challenge to investigate the reproducibility property both theoretically and experimentally. Thereby, a very small magnetic field dependent quantum electrodynamics correction to $R_\mathrm{K}$ was recently predicted\cite{Penin2009}. Although its measurement is far beyond the present time capability, any new experimental test of the reproducibility property performed with a better accuracy will reinforce our confidence in the QHE as a conerstone of the future \emph{Syst\`{e}me International} of units based on the fixing of fundamental constants of physics\cite{Mills2006}. A major stake is the redefinition of the kilogram in terms of \emph{h} by means of the QHE and JE in the watt balance experiment\cite{Eichenberger2009}. Any lack of reproducibility of the QHE will further question the relation $R_\mathrm{K}=h/e^2$ and impact the \emph{SI} redefinition.

The most severe reproducibility tests of the QHE have been performed by comparing QHR in different materials leading hence to the so-called universality tests. QHR realized in GaAs/AlGaAs 2DEG and Si-Mosfet based samples\cite{Hartland1991,Jeckelmann1997} were found to agree with a relative uncertainty of $3\times 10^{-10}$. More recently, graphene has offered the opportunity of new stringent tests\cite{Poirier2010}: the QHR on the $\nu=2$ plateau in graphene monolayer\cite{Tzalenchuk2010, Janssen2011} (resp. $\nu=-4$ plateau in graphene bilayer\cite{Guignard2012}) was recently found in agreement with the QHR in GaAs/AlGaAs within a relative uncertainty of $9\times 10^{-11}$ (resp. $5\times 10^{-7}$). Further reducing these uncertainties is timely to deepen the investigation of the relativistic QHE in graphene. The comparison between the integer and fractional QHE is another way to test the universality property\cite{Ahlers2008}. More precise reproducibility tests can be performed by comparing quantum standards made of the same material and varying other experimental implementation conditions.

In this paper, we report on a comparison, performed using the original quantum Wheatstone bridge technique\cite{Schopfer2007}, of four different QHR, each made of the same GaAs/AlGaAs 2DEG widely used in metrology institutes\cite{Piquemal1993}. A negligible relative deviation of $-1.9\times 10^{-12}$ covered by a record measurement relative uncertainty of $32\times 10^{-12}$ ($1\sigma$) is found between the QHR in the limit of dissipationless state. This new demonstration of reproducibility has a relative uncertainty three order lower than that of the $R_\mathrm{K}$ determination in the \emph{SI} using the Thompson-Lampard capacitor\cite{Poirier2011}. This uncertainty is also three times smaller than that of the best QHR comparison between graphene and GaAs \cite{Janssen2011}. Moreover, the measurement of the discrepancy between QHR and of the longitudinal resistance as a function of magnetic field very close to the zero-dissipation state shows a resistivity rule between the diagonal ($R_{xx}$) and off-diagonal components of the resistivity tensor ($R_{xy}=R_\mathrm{H}$) . These results not only consolidate the QHE physics but also bring deepen knowledge for improving the practical realization of the resistance unit.

\section{\label{sec:level1}Experimental setup and measurement technique}
Fig.~1 shows the Wheatstone bridge device made of four $400~\mu\mathrm{m}$-wide Hall bars connected by aluminum bonding wires using the triple connection technique. Exploiting the fundamental properties of the QHE, this method\cite{Delahaye1993} reduces the impact of the wire and contact resistance $R_\mathrm{C}$ on the four-terminal definition of the Hall resistance $R_\mathrm{H}$ to a third order effect $O[(R_\mathrm{C}/R_\mathrm{H})^3]$. Fig. 2 shows the typical magneto-resistance curves of one Hall bar measured at 1.3 K with a current of $10~\mu A$. The four samples are characterized by very close electronic mobility ($310000~\mathrm{cm}^2 \mathrm{V}^{-1}\mathrm{s}^{-1}$) and carrier density ($5.2\times 10^{11}~\mathrm{cm}^{-2}$) values. Each Hall bar has been individually checked according to the technical guidelines for the QHE metrological use\cite{Delahaye2003}. Notably, the resistance of each AuGeNi ohmic contact\cite{Jeckelmann2001} was deduced from a three-terminal resistance measurement in the QHE regime at $\nu=2$. In such a configuration, the contact of interest is used to inject the current and probe the potential, a second contact is used to expel the current and a third contact on the equipotential where lies the contact of interest is used to probe the voltage drop, so that the measured quantity is the series of the cable resistance between the top of the refrigerator and the contact of interest, the contact resistance itself and the longitudinal resistance. Knowing the cable resistance ($0.3~\Omega$), and the longitudinal resistance along the equipotential at $\nu=2$ in these samples being negligible ($<100~\mu\Omega$), the resistance of each contact can be easily deduced to be lower than $0.1~\Omega$. Once the Wheatstone bridge mounted, using a similar three-terminal measurement technique applied at $\nu=2$, the resistance of the twelve accessible terminals $Inj_\mathrm{A}^{+}$, $L_\mathrm{A}^{+}$, $H_\mathrm{A}^{+}$, $Inj_\mathrm{B}^{+}$, $L_\mathrm{B}^{+}$, $H_\mathrm{B}^{+}$, $Inj_\mathrm{A}^{-}$, $L_\mathrm{A}^{-}$, $H_\mathrm{A}^{-}$, $Inj_\mathrm{B}^{-}$, $L_\mathrm{B}^{-}$, $H_\mathrm{B}^{-}$ (see Fig.~1) was determined. For example, in the case where the current circulates between $Inj_\mathrm{A}^{+}$ and $Inj_\mathrm{A}^{-}$ and the voltage drop is measured between $Inj_\mathrm{A}^{+}$ and $H_\mathrm{A}^{+}$, the resistance that can be deduced is half the mean value of the resistance of current contact of Hall bar 1 and the resistance of current contact of Hall bar 4. From these twelve measurements it was also deduced that the mean value of the contact resistance is below $0.1~\Omega$. The resistance comparison error due to interconnections is therefore strongly reduced by the multiple connection technique to a negligible level of $\sim 10^{-15}$. All measurements presented in the following were carried out at magnetic inductions corresponding to the $R_\mathrm{K}/2$ plateau for each Hall bar.
\begin{figure}[h]
\begin{center}
\includegraphics[width=8.5cm]{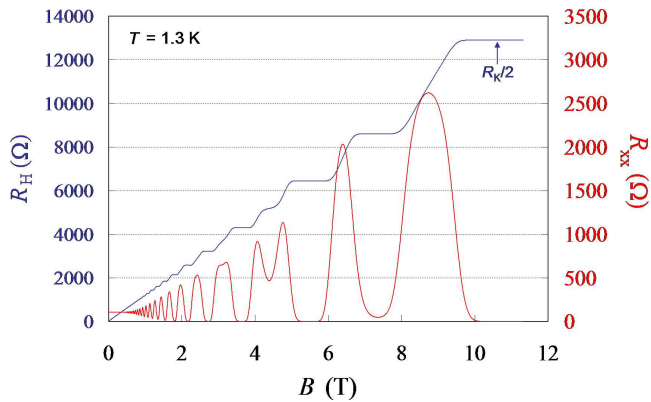}
\caption{Hall resistance and longitudinal resistivity measured with a current of $10~\mu A$ at $T$=1.3 K as a function of the magnetic induction in Hall bar sample numbered 2}\label{fig2}
\end{center}
\end{figure}
Fig.~1 presents the drawing of the Wheatstone bridge integrated into the experimental setup. The connection scheme enables the measurement of the resistance unbalance of the bridge in two configurations $\mathrm{C_A}$ and $\mathrm{C_B}$ along the two diagonals of the device. In configuration $\mathrm{C_A}$, it has been measured using the supplying terminal-pair ($Inj_\mathrm{A}^{+}$, $Inj_\mathrm{A}^{-}$) and current detection terminal-pairs ($W_\mathrm{A}^{+}$, $W_\mathrm{A}^{-}$) and in configuration $\mathrm{C_B}$ using the terminal-pairs ($Inj_\mathrm{B}^{+}$, $Inj_\mathrm{B}^{-}$) for the supply and ($W_\mathrm{B}^{+}$, $W_\mathrm{B}^{-}$) for the detection. In the Direct Current (DC) limit the relative unbalance current $I_{ub}/I$ of the Wheatstone bridge, where $I$ is the current circulating in each Hall bar, is related to the relative deviations of each QHR $j$ to $R_\mathrm{K}/2$, $\alpha_j=(2R_\mathrm{H}^j-R_\mathrm{K})/R_\mathrm{K}$:
\begin{displaymath}
[I_{ub}/I]_\mathrm{C_A}=-[I_{ub}/I]_\mathrm{C_B}=\frac{[(\alpha_2+\alpha_4)-(\alpha_1+\alpha_3)]}{2}
\end{displaymath}
It can be interpreted as the relative deviation of one particular quantum resistor among the others:
\begin{displaymath}
\Delta R/R= 2[I_{ub}/I]_\mathrm{C_A}=-2[I_{ub}/I]_\mathrm{C_B}
\end{displaymath}
 In our experiment, the Wheatstone bridge is biased by a low frequency $f$ alternating voltage (between 0.10 Hz and 20 Hz) and the unbalance current ${I_{ub}}(f)$ is detected by a 3436 turns winding of a CCC used as amplifier. The $13.6~\mu\mathrm{A.turns/\phi_0}$ sensitivity of the CCC equipped with a $\mathrm{10^{-4}~\phi_0/Hz^{1/2}}$ white noise radio-frequency SQUID results in a $\mathrm{400~fA/Hz^{1/2}}$ current resolution. The polarization of the bridge and the signal detection at the output of the SQUID electronics are ensured by a Signal Recovery 7265 lock-in detector. At each measurement frequency, the angular phase of the lock-in detector is adjusted with a standard uncertainty of $20~\mu \mathrm{rad}$ so that placing a $1~\mathrm{M\Omega}$ resistor parallel to one arm of the bridge, which produces a real impedance (resistance) unbalance, results in a calibrated signal only on the in-phase lock-in axis. Afterwards, the in-phase signal at frequency $f$ can be converted to an equivalent resistance unbalance $(\Delta R/R)_f$. The latter includes the QHR unbalance $\Delta R/R$ and others very small residual frequency dependent contributions caused by the measurement chain. The measurement procedure then consists in determining precisely the frequency dependence of $(\Delta R/R)_f$ to obtain $\Delta R/R$ alone by extrapolation to zero frequency. Each cable connected to the bridge is PTFE insulated and protected by a shielding at ground potential. Thus, only current leakage to ground can affect the comparison accuracy. But the symmetric polarization of the bridge sets current detection terminals at potential very close to ground: in-phase (out-phase) voltage is no more than $10^{-5}$ ($10^{-4}$) times $R_\mathrm{H}I$. It results that the measurement error of $\Delta R/R$ caused by current leakage is strongly reduced to about $10^{-5}\times R_\mathrm{H}/R_\mathrm{G}$, where $R_\mathrm{G}\gg10^{12}~\Omega$ is the insulation resistance between cable inner and ground. The measurement uncertainty of $\Delta R/R$ is therefore determined from the statistical analysis ($1\sigma$) of the measurement noise (type A uncertainty) as reported in all figures.
\section{\label{sec:level1}Results}
\subsection{\label{sec:level2}Measurement of the longitudinal resistance}
\begin{figure}[h]
\begin{center}
\includegraphics[width=8.5cm]{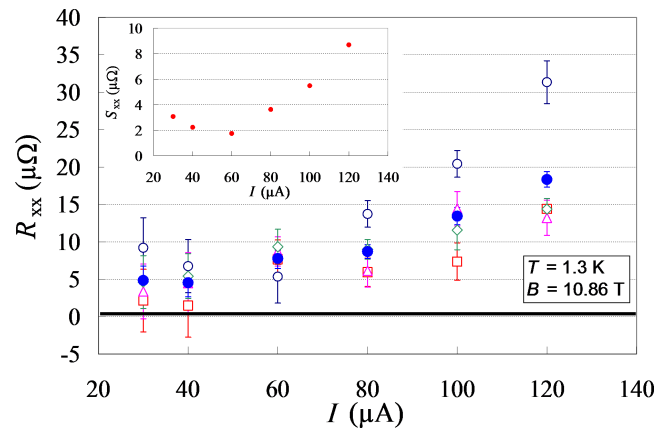}
\caption{Current dependence of $R_{xx}(\mathrm{A^{+}})$ (open blue circle), $R_{xx}(\mathrm{A^{-}})$ (open red square), $R_{xx}(\mathrm{B^{+}})$ (open magenta triangle) , $R_{xx}(\mathrm{B^{-}})$ (open green diamond), mean value $\overline{R_{xx}}$ (filled blue circle). Inset: standard deviation $S_{xx}$.}\label{fig3}
\end{center}
\end{figure}
The perfect quantization state of the QHE is only reached in the zero-dissipation limit. The longitudinal resistivity which is assumed to measure the dissipation level cannot be determined independently in each quantum resistance standard once the Wheatstone bridge is mounted. But, using $H_\mathrm{A}^{+}$,$H_\mathrm{A}^{-}$,$L_\mathrm{A}^{+}$,$L_\mathrm{A}^{-}$ voltage terminals, four resistances which are different combination of two longitudinal resistivity $r_{xx}$ of individual Hall bars were measured using a EMN11 nanovoltmeter with an accuracy around $1~\mu\mathrm{\Omega}$:
\begin{eqnarray*}
R_{xx}(\mathrm{A}^{+})&=&\frac{W}{L}\frac{|V(H_\mathrm{A}^{+})-V(L_\mathrm{A}^{+})|}{I}=\frac{(r_{xx}^{1,Up}+r_{xx}^{4,Up})}{2} \\
R_{xx}(\mathrm{A}^{-})&=&\frac{W}{L}\frac{|V(H_\mathrm{A}^{-})-V(L_\mathrm{A}^{-})|}{I}=\frac{(r_{xx}^{2,Lo}+r_{xx}^{3,Lo})}{2} \\
R_{xx}(\mathrm{B}^{+})&=&\frac{W}{L}\frac{|V(H_\mathrm{B}^{+})-V(L_\mathrm{B}^{+})|}{I}=\frac{(r_{xx}^{2,Up}+r_{xx}^{1,Lo})}{2} \\
R_{xx}(\mathrm{B}^{-})&=&\frac{W}{L}\frac{|V(H_\mathrm{B}^{-})-V(H_\mathrm{B}^{-})|}{I}=\frac{(r_{xx}^{4,Lo}+r_{xx}^{3,Up})}{2}
\end{eqnarray*}
(Up and Lo exponents which complete the $r_{xx}^{j}$ notation refers to upper and lower edges of the Hall bar $j$ respectively).
Fig. 3 shows that the current behaviors of these four $R_{xx}$ quantities, measured at the center of the $\nu=2$ plateau (\emph{B}=10.86 T), are similar in the current range investigated. Below $60~\mu A$, all longitudinal resistances are independent of the current value and their values, only determined by the temperature $T$=1.3 K, represents no more than one part in $10^9$ of the Hall resistance. The current behavior of their mean value $\overline{R_{xx}}$, equal to the mean value $\overline{r_{xx}}$, seems therefore very representative of the behavior with current in each Hall bar. In addition, $R_{xx}(\mathrm{A}^{+})$ discards much more from the mean value $\overline{R_{xx}}$. This is in agreement with the conclusion drawn from individual characterizations showing that $r_{xx}^{1,Up}$ was higher in the Hall bar sample numbered 1. Although the resistances $R_{xx}(\mathrm{A}^{+})$, $R_{xx}(\mathrm{A}^{-})$, $R_{xx}(\mathrm{B}^{+})$ and $R_{xx}(\mathrm{B}^{-})$ are correlated because $r_{xx}^{j,Up}$ and $r_{xx}^{j,Lo}$ characterize the same sample $j$, the standard deviation $S_{xx}$\cite{Sxxdefinition} is a useful quantity reflecting the $r_{xx}$ dispersion among Hall bars. Its dependence on current is similar to that of $\overline{R_{xx}}$ (see inset of fig.3). At this point, we want to make clear that the identification of the mechanisms leading to the dissipative transport in the QHE regime is beyond the scope of this paper. Even if it is expected that variable range hopping \cite{Shklovskii1984} could explain $r_{xx}$ at low temperature and low current\cite{Jeckelmann2001}, it would have been useful to this end to study the current dependence or, better, the temperature dependence of the longitudinal resistivity.

\subsection{\label{sec:level2}Hall resistance plateau tilting and the resistivity rule}
\begin{figure}[h]
\begin{center}
\includegraphics[width=8.5cm]{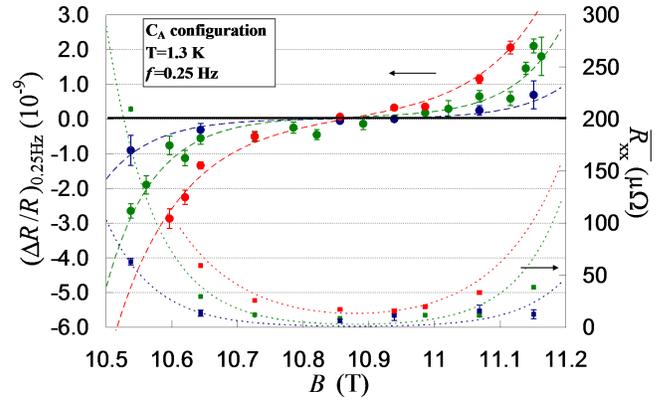}
\caption{$(\Delta R/R)_{0.25 Hz}$ and $\overline{R_{xx}}$ as a function of the magnetic induction \emph{B} for $40~\mu A$ (blue), $80~\mu A$ (green) and $120~\mu A$ (red) measurement currents. Dashed lines are interpolation curves of $(\Delta R/R)_{0.25 Hz}$. Dotted lines are representations of $2.7\times 10^{-2}\times B\times(d(\Delta R)/d B)_{0.25 Hz}$ functions.}\label{fig4}
\end{center}
\end{figure}
Fig. 4 shows that the discrepancy $(\Delta R/R)_{0.25 Hz}$, measured in configuration $\mathrm{C_A}$ at $T = 1.3~\mathrm{K}$, as a function of the magnetic induction \emph{B} forms a plateau with a finite slope increasing with current. All plateaus measured with different currents intercept zero discrepancy value at a common magnetic induction $B_\mathrm{P}=10.86~\mathrm{T}$ defining an apparently current independent fixed pivotal point within a $10^{-10}$ uncertainty. At $B_\mathrm{P}$, $\overline{R_{xx}}$ is minimal and the Landau level filling factor $\nu=\overline{n_{\mathrm{s}}}h/(eB_\mathrm{P})$, calculated using the mean value of the electron densities in the four Hall bars  $\overline{n_{\mathrm{s}}}=5.19\times10^{11}~\mathrm{cm^{-2}}$ is 1.98, therefore very close to 2. The observation of such a fixed point comes from the electron densities of the four Hall bars differing between them by less than $10^{-2}$ in relative value. $\overline{R_{xx}(B)}$ at each current can be very well adjusted by the function $2.7\times 10^{-2}\times B\times(d(\Delta R(B))/d B)_{0.25 Hz}$ (see dotted lines in figure 4), with $(d(\Delta R(B))/d B)_{0.25 Hz}$ the derivative of the functions interpolating $(\Delta R/R)_{0.25 Hz}$ curves. The linear relationship between the slope $(d(\Delta R(B))/d B)_{0.25 Hz}$ and $\overline{R_{xx}(B)}$ notably means that the plateau is perfectly flat in the dissipation-less limit, as expected. Similarly, it is found that $S_{xx}(B)\sim 2.3\times 10^{-2}\times B\times(d(\Delta R(B))/d B)_{0.25 Hz}$ showing that the observation of $(\Delta R/R)_{0.25 Hz}$ plateau tilting in our experiment probably results from the small difference existing between $r_{xx}$ values in the four Hall bars as yet concluded from fig. 2. The observation of a linear relationship between the Hall plateau slope and the longitudinal resistance, know as the resistivity rule (RR)\cite{Chang1985, Rotger1989, Stormer1992, Pan2005}, was reported in 2DEG by several groups. This is a very general relation which also applies to the thermopower tensor in the QHE regime\cite{Tieke1997}. It was proposed that this linear relationship results from large scale density fluctuations\cite{Simon1994,Simon1997}. In this model, the macroscopic dissipative resistivity $R_{xx}$ depends mainly on the fluctuations of the microscopic transverse resistivity $\delta\rho_{xy}$ caused by the spatial variations of the filling factor and is weakly linked to the microscopic resistivity $\rho_{xx}$ assumed to be $\ll\delta\rho_{xy}$. It results that $R_{xx}(B)\sim \alpha\times B\times dR_{xy}(B)/d B$ where $\alpha$ is given by $C\delta n_\mathrm{s}/n_{\mathrm{s}}$ and $C$ is assumed to be order unity. The value of $\alpha=2.7\times 10^{-2}$ deduced from our adjustments is in agreement with the relative density fluctuations of $10^{-2}$ measured between samples. The observation of the (RR) on the $\nu=2$ plateau at the $10^{-9}$ accuracy seems to indicate that the microscopic dissipation characterized by $\rho_{xx}$ is very low. In the scope of this model, $\overline{R_{xx}}$ would simply be an upper bound of $\rho_{xx}$, which turns out to be very low. However, the longitudinal resistance stays a relevant quantization criterion since directly linked to Hall resistance deviations. More practically, it appears that reproducing $R_{\mathrm{K}}$ value within $10^{-10}$ requires samples with very low $r_{xx}$ value and highly homogeneous carrier density as well as very precise determination of the magnetic field giving the minimum of $r_{xx}$.
\subsection{\label{sec:level2}Frequency dependence of the Hall resistance discrepancy}
At $B_\mathrm{P}=10.86~\mathrm{T}$, the $\nu=2$ plateau center which gives minimal discrepancy as well as minimal $\overline{R_{xx}}$ value, the frequency dependence of $(\Delta R/R)_{f}$ between 0.15 Hz and 20 Hz was accurately measured
 to determine $\Delta R/R$ by extrapolation to $f=0~\mathrm{Hz}$. These measurements were performed for current values of $40~\mathrm{\mu A}$, $80~\mathrm{\mu A}$ and $120~\mathrm{\mu A}$ using measurement configurations $\mathrm{C_A}$ and $\mathrm{C_B}$.
  Fig. 5 reports on $(\Delta R/R)_{f}$ ($Im(\Delta Z/Z)_f$, the imaginary part of the impedance unbalance, in inset) as a function of frequency $f$ for measurement configuration $\mathrm{C_A}$ and a current of $80~\mathrm{\mu A}$. Each determination at a given frequency is the weighted mean value of several measurements with acquisition time as long as 50~000 s. Allan variance analysis shows that the standard mean deviation is an adequate estimate of the uncertainty (noise is white, no $1/f$ noise is observable). $(\Delta R/R)_{f}$ can be perfectly adjusted by a second order polynomial function which is different for $\mathrm{C_A}$ or $\mathrm{C_B}$ measurement configurations. The understanding of the frequency dependence is beyond the scope of the paper. However, let us just remark that the linear frequency dependent term has a significant amplitude of $-3.07\times 10^{-7}$/kHz in configuration $\mathrm{C_A}$ that could result from losses in capacitances as was observed at higher frequency (a few kHz)\cite{Ahlers2009}.
\begin{figure}[h]
\begin{center}
\includegraphics[width=8.5cm]{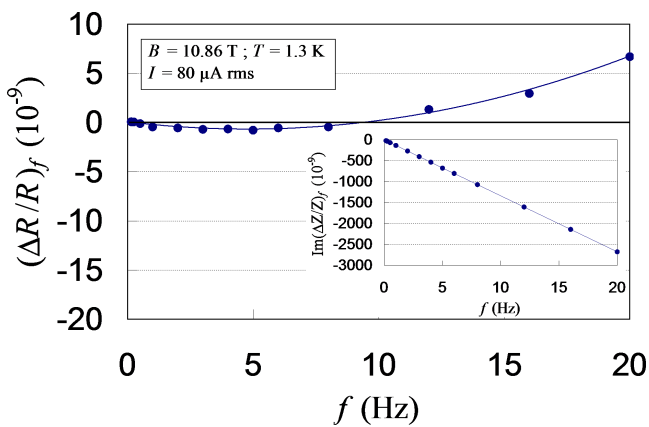}
\caption{Frequency dependence of $(\Delta R/R)_{f}$ ($Im(\Delta Z/Z)_f$ in inset) in measurement configuration $\mathrm{C_A}$. Uncertainty bars are smaller than blue circle size.}\label{fig5}
\end{center}
\end{figure}
\subsection{\label{sec:level2}Reproducibility of the Hall resistance close to the zero-dissipation state}
\begin{table}
\caption{\label{tab:table 1}Current dependence of mean longitudinal resistance and relative deviations to quantization.}
\begin{ruledtabular}
\begin{tabular}{ccccc}
\mbox{$I$}&\mbox{$\overline{R_{xx}}$}&\mbox{$(\Delta R/R)_{\mathrm{C_A}}$}&\mbox{$(\Delta R/R)_{\mathrm{C_B}}$}&\mbox{$(\Delta R/R)_{\mathrm {C_{AB}}}$}\\
\mbox{$(\mu A)$}&\mbox{$(\mu \Omega)$}&\mbox{$(10^{-12})$}&\mbox{$(10^{-12})$}&\mbox{$(10^{-12})$}\\
\hline
40&$6.2\pm1.6$&$-3.9\pm48.7$&$17.6\pm71.3$&$2.9\pm40.2$\\
80&$8.5\pm0.8$&$68.4\pm20.6$&$48.3\pm19.5$&$57.8\pm14.2$\\
120&$17.1\pm0.9$&$119.4\pm19.3$&$48.6\pm37.5$&$104.5\pm17.2$\\
\end{tabular}
\end{ruledtabular}
\end{table}
The extrapolation of $(\Delta R/R)_{f}$ to $f=0$ Hz measured in configurations $\mathrm{C_A}$ and $\mathrm{C_B}$ for each current results in $\Delta R/R_{\mathrm{C_A}}(I)$ and $\Delta R/R_{\mathrm{C_B}}(I)$ respectively, the weighted mean value $\Delta R/R_{\mathrm{C_{AB}}}(I)$ of which can then be calculated. All values are reported in Table 1. Inset of Fig.6 shows $\Delta R/R_{\mathrm{C_{AB}}}(I)$ as a function of current. $\Delta R/R_{\mathrm{C_{AB}}}(I)$ reported as a function of $\overline{R_{xx}}(I)$ (see fig. 6) exhibits a linear relationship with a coupling factor of 0.08. We then used a generalized $\chi$-two linear least square method to extrapolate $\Delta R/R_{\mathrm{C_{AB}}}$ to $\overline{R_{xx}}=0$. The result is $\Delta R/R_{\mathrm{C_{AB}}}(\overline{R_{xx}}=0)= (-1.9 \pm 31.8)\times 10^{-12}$. It demonstrates with a record relative uncertainty of $32\times 10^{-12}$ the reproducibility of the quantized Hall resistance on the $\nu=2$ plateau in the GaAs samples if they are in the dissipation-less state. For a minimal $\overline{R_{xx}}$ value higher than $20~\mu\Omega$, the QHR departs from each others by more than $1\times 10^{-10}$ at the center of the $\nu=2$ plateau.
 $\Delta R/R_{\mathrm{C_{AB}}}(I)$ was also reported as a function of $S_{xx}(I)$ showing a linear relationship. The extrapolation to $S_{xx}=0$ gives $\Delta R/R_{\mathrm{C_{AB}}}(S_{xx}=0)=(12 \pm 29.9)\times 10^{-12}$. This result means that the discrepancy between quantum Hall resistances cancels if the four Hall bars have the same macroscopic longitudinal resistance. This claims for a similar mechanism linking the Hall and the longitudinal resistances in each Hall bar. This confirms the hypothesis previously evoked that the discrepancy observed at finite current probably results from unequal longitudinal resistance values in the four Hall bars.
 \begin{figure}[h]
\begin{center}
\includegraphics[width=8.5cm]{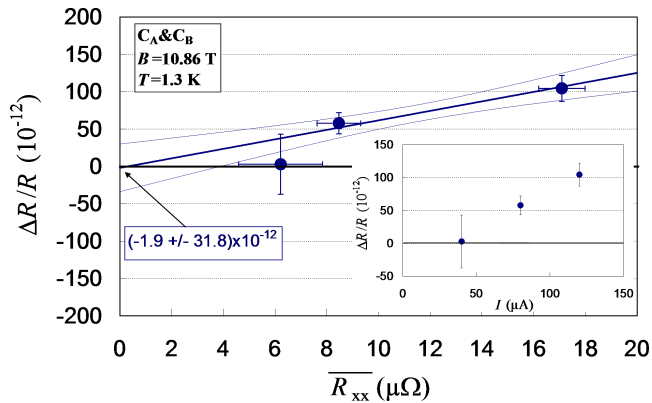}
\caption{Relative resistance deviation between quantum Hall resistances $\Delta R/R_{\mathrm{C_{AB}}}$ as a function of the mean longitudinal resistance $\overline{R_{xx}}$. Linear adjustment of $\Delta R/R_{\mathrm{C_{AB}}}(\overline{R_{xx}})$ (thick blue line) and $1\sigma$ standard deviations (thin blue line). Inset: $\Delta R/R_{\mathrm{C_{AB}}}$ as a function of current.}\label{fig6}
\end{center}
\end{figure}
 Considering the magnetic field dependence of $(\Delta R/R)_{0.25 Hz}$ expressed by the (RR)(see fig. 3), the linear behavior of $\Delta R/R$ as a function of $\overline{R_{xx}}$ can simply result either from the $\sim 10^{-2}$ T-inaccuracy of the magnetic induction experimental realization of the pivotal point or from its indetermination itself due to the $10^{-2}$-dispersion of the electron densities of the four Hall bars. The existence of small effective misalignments of QHR Hall probes, usually explained by the carrier density inhomogeneity\cite{DominguezPhd1988} or the chiral nature of the current flow in finite width voltage terminals\cite{Vanderwel1988}, can also explain the observed linear coupling. Although samples have very close geometry, the deviations caused by the finite width of voltage terminals should not totally compensate since longitudinal resistances are different in the four Hall bars.
\section{\label{sec:level1}Conclusion}
This work demonstrates the perfect reproducibility of the quantum Hall effect with a record uncertainty of $32\times 10^{-12}$. This is a new result supporting a revision towards a \emph{Syst\`{e}me International} of units taking into account quantum physics. It also shows that, close to the zero dissipation state, the Hall resistance on the $\nu=2$ plateau departs from $R_\mathrm{K}/2$ as a function of minimal longitudinal resistivity following the so-called resistivity rule. It turns out that a longitudinal resistance higher than $20~\mu\Omega$ leads to a relative discrepancy to quantization of the Hall resistance higher than $10^{-10}$. The accuracy of the reported comparison was limited by the sensitivity of the CCC. A current sensitivity of $\mathrm{20~fA/Hz^{1/2}}$ can be achieved using a 4000 turns detection winding of a DC SQUID based CCC\cite{Soukiassian2010} with a sensitivity of about $\mathrm{80~pA.turn/Hz^{1/2}}$. Comparisons of quantum resistance standards, for example made of GaAs and graphene\cite{Poirier2010}, could therefore be performed with an accuracy $\sim 10^{-12}$ using the Wheatstone QHE bridge technique.

\smallskip
This research has received funding from the European Community's Seventh Framework Programme, ERA-NET Plus, under Grant Agreement No. 217257.

\providecommand{\noopsort}[1]{}\providecommand{\singleletter}[1]{#1}%

\end{document}